# The genomic impacts of drift and selection for hybrid performance in maize

Justin P. Gerke[*,1], Jode W. Edwards[§], Katherine E. Guill[†], Jeffrey Ross-Ibarra[‡] and Michael D. McMullen[*,†]

[*]Division of Plant Sciences, University of Missouri, Columbia MO 65211

[§]Corn Insects and Crop Genetics Research Unit, USDA-Agricultural Research Service, Ames, IA, 50011

[†]Plant Genetics Research Unit, USDA-Agricultural Research Service, Columbia MO 65211

[‡]Department of Plant Sciences, Center for Population Biology, and Genome Center, University of California, Davis, CA 95616

---

[1] Present address: DuPont Pioneer, Johnston, Iowa

**Corresponding Author:**
Justin P. Gerke
DuPont Pioneer
8305 NW 62$^{ND}$ Ave
P.O. Box 7060
Johnston, IA, 50131
justin.gerke@gmail.com

ABSTRACT

Modern maize breeding relies upon selection in inbreeding populations to improve the performance of cross-population hybrids. The United States Department of Agriculture - Agricultural Research Service (USDA-ARS) reciprocal recurrent selection experiment between the Iowa Stiff Stalk Synthetic (BSSS) and the Iowa Corn Borer Synthetic No. 1 (BSCB1) populations represents one of the longest standing models of selection for hybrid performance. To investigate the genomic impact of this selection program, we used the Illumina MaizeSNP50 high-density SNP array to determine genotypes of progenitor lines and over 600 individuals across multiple cycles of selection. Consistent with previous research (Messmer et al., 1991; Labate et al., 1997; Hagdorn et al., 2003; Hinze et al., 2005), we found that genetic diversity within each population steadily decreases, with a corresponding increase in population structure. High marker density also enabled the first view of haplotype ancestry, fixation and recombination within this historic maize experiment. Extensive regions of haplotype fixation within each population are visible in the pericentromeric regions, where large blocks trace back to single founder inbreds. Simulation attributes most of the observed reduction in genetic diversity to genetic drift. Signatures of selection were difficult to observe in the background of this strong genetic drift, but heterozygosity in each population has fallen more than expected. Regions of haplotype fixation represent the most likely targets of selection, but as observed in other germplasm selected for hybrid performance (Feng et al., 2006), there is no overlap between the most likely targets of selection in the two populations. We discuss how this pattern is likely to occur during selection for hybrid performance, and how it poses challenges for dissecting the impacts of modern breeding and selection on the maize genome.

INTRODUCTION

Although maize is naturally an out-crossing organism, modern breeding develops highly inbred lines that are then used in controlled crosses to produce hybrids. Hybrid maize has been so successful that it quickly replaced long-standing mass-selected open pollinated varieties (Crabb, 1947). Maize inbred lines are now partitioned into separate inbreeding 'heterotic groups' that maximize performance and hybrid vigor (heterosis) when an inbred line of one group is crossed with the other group (Tracy and Chandler, 2006). The shift towards selection of inbred lines based on their ability to generate good hybrids – referred to as 'combining ability' – constituted an abrupt change from the open-pollinated mass selection that breeders practiced for millennia (Anderson, 1944; Troyer, 1999).

Multiple studies with molecular markers have indicated that the modern era of single-cross hybrid maize breeding has led to a dramatic restructuring of population genetic variation (Duvick et al., 2004; Ho et al., 2005; Feng et al., 2006). Different heterotic groups have diverged genetically over time to become highly structured and isolated populations. Advances in high throughput genotyping and the development of a maize reference genome now enable the observation of maize population structure at high marker density across the whole genome (Ganal et al., 2011; Chia et al., 2012). So far, these high-density studies have examined a broad spectrum of germplasm at various points in the history of maize to search for the signals of population structure and artificial selection (Hufford et al., 2012; van Heerwaarden et al., 2012). Although selective sweeps remaining from domestication are clearly visible, the impact of selection during modern breeding appears comparatively small in terms of its impact on genomic diversity despite steady, heritable improvement in phenotype (Duvick, 2005). The lack of distinct selection signals from modern breeding may be due to specific selective events occurring in different populations, necessitating a more focused look within heterotic groups or even single breeding programs.

In this study, we apply a genomic approach to study the dynamics of genetic variation over time within an individual selection experiment. The Iowa Stiff Stalk Synthetic (BSSS) and the Iowa Corn Borer Synthetic No. 1 (BSCB1) Reciprocal Recurrent Selection Program of the USDA-ARS at Ames, Iowa (hereafter referred to as the Iowa RRS) represents one of the best-documented public experiments on selection for combining ability and hybrid performance. BSSS and BSCB1 have been recurrently selected for improved cross-population hybrids (Penny and Eberhart, 1971). This model of selection, named reciprocal recurrent selection, provides the generalized model for strategies used in commercial maize breeding (Comstock et al., 1949; Duvick et al., 2004). The Iowa RRS experiment proves especially relevant because lines derived from the BSSS population have had a major impact upon the development of commercial hybrids (Duvick et al., 2004; Darrah and Zuber, 1986), the formation of modern heterotic groups (Troyer, 1999; Senior et al., 1998), and the choice of a maize reference genome (Schnable et al., 2009).

## MATERIALS AND METHODS

**The BSSS and BSCB1 Recurrent Selection Program:** The Iowa RRS experiment began with founder inbred lines (Table S1) that were randomly mated to create the BSSS and BSCB1 'cycle 0' base populations. Each group was then recurrently selected for improved combining ability with the other group (Penny and Eberhart, 1971; Keeratinijakal and Lamkey, 1993). Initially approximately 200-250 starting 'S0' plants within each population were simultaneously self-fertilized to generate 'S1' lines and crossed to a random sample (4-6) of plants from the other population to generate testcross seed. The material was evaluated for highly heritable traits including standability, disease resistance and ear rot resistance in the nursery at the time the testcross seed was made, and 100 testcrosses were then grown in replicated field trials and evaluated for grain yield, grain moisture, and standability. S1 seed of 10 selected families within each population were randomly mated to create a new 'cycle n+1' population. After cycle 4, the testcrosses were carried out on the S1 lines themselves rather than at S0, which leads to another round of selfing prior to the creation of the next cycle. Ten lines (out of 100) were selected and advanced to form the next cycle between cycles 1 and 8, and twenty lines were selected between cycles 9 and 16 (Keeratinijakal and Lamkey, 1993). Founder inbreds and samples from cycles 0, 4, 8, 12, and 16 were genotyped at 39,258 SNPs that passed a set of quality filters and could be assigned collinear genetic and physical map positions.

**Plants and inbred lines used:** The plants and inbred lines used in this experiment are listed in Table S1. We genotyped 36 plants from each of the BSSS and BSCB1 populations at selection cycles 0, 4, 8, 12, and 16. These plants represent descendants of the original populations which have been randomly mated to maintain seed. We also genotyped the founder inbreds for each population, however, seed for the founder lines F1B1, CI.617, WD456, and K230 were not available. The data for founder line CI.540 was not used because the genotyped material was heterozygous. A number of derived lines were also genotyped for calibrating phasing and imputation procedures (see below).

**Genotype data:** Plants from the cycles of selection, founders, and derived lines were grown in a greenhouse and tissue was collected at the 3 leaf stage. Tissue was lyophilized, ground, and DNA extracted by a CTAB procedure (Saghai-Maroof et al., 1984). Samples were genotyped using the 24-sample Illumina MaizeSNP50 array (Ganal et al., 2011) according to the Illumina Infinium protocol, and imaged on an Illumina BeadStation at the University of Missouri DNA core facility. Genotypes were determined with the GenomeStudio v2010.2 software using the MaizeSNP50_B.egt cluster file, (http://www.illumina.com/support/downloads.ilmn). The design of the maize SNP50 Chip included a relatively small ascertainment panel of inbred lines, introducing a bias in the frequencies of SNPs included on the chip (Ganal et al., 2011). However, because our simulations are based not on theoretical expectations but instead on sampling from the observed data at cycle 0, we expect ascertainment bias to have a minimal impact on our results. The effect of this ascertainment bias was shown to be minimal in another study of North American germplasm (van Heerwaarden et al., 2012).

48,919 SNPs were called on the Illumina platform from the MaizeSNP50_B.egt cluster file. Genotypes with quality scores of 50 or less were recoded as missing data. Three plants were removed from the data due to an excess of missing data (the derived line B10, a cycle 0 plant from BSSS, and a cycle 8 plant from BSSS). In addition, BSCB1 plant 31 from cycle 4 appeared switched with plant 31 from cycle 8 based on our principal component analysis (PCA), so we switched the labels for these two genotypes to correct the mistake. To avoid structure among the missing data, we removed any SNP that was coded as missing in more than 3 plants in either group of founders or any group of plants from a particular cycle and population. Preliminary analysis by PCA and heatmap plots of distance matrices revealed two additional likely mix-ups.

Plant 23 from BSSS cycle 8 was a clear outlier from the BSSS population as a whole and plant 2 from BSSS cycle 0 is likely a mislabeled plant from cycle 16. Since there was no evidence suggesting when mis-labelings occurred, each of these plants was removed from the analysis.

**Integrating the genetic and physical map:** The interpretation of our results depends upon a genetic and physical map that is as accurate as possible. We therefore took steps to improve the positions of the SNP markers on the genetic and physical maps relative to version 5A.59 of the maize genome assembly (maizesequence.org). The probable physical position of each SNP based was obtained by comparing SNP context sequences to the genome sequence. For this purpose, SNP context sequences were defined as the sequence 25 bp upstream of the SNP, the bp representing the SNP itself, followed by 25 bp downstream of the SNP, making a total sequence length of 51bp. When a single genomic location was queried by two separate probes on the array, we chose the probe with higher quality calls and dropped the other marker from the dataset.

To assign a genetic position for each SNP, we used a map derived from the B73xMo17 (IBM) mapping population similar to the IBM framework map in Ganal et al. (Ganal et al., 2011). This genetic map contains 4,217 framework SNP markers, which provides a much higher density than the map used to order the 5A.59 release of the maize genome sequence. As a result, we identified several places in the genome where the physical positions were incorrect according to our genetic map. These cases included both simple reversals of the physical map relative to the genetic map, and also the assignment of blocks of markers to the wrong linkage group, which we refer to as mis-mapped blocks. SNPs at these loci were reordered to match our genetic map. To maintain collinearity between the genetic and physical map, the physical positions of these SNPs were reassigned as follows. Individual mis-mapped markers were simply removed from the data. We also removed small reversals and mis-mapped blocks (<10 kb). Small rearrangements of this sort are more likely to represent mis-mapped paralagous sequence than true errors in the physical map. When larger reversals were identified, we transposed the physical positions of the SNPs from one end of the segment to the other. Mis-mapped blocks were often larger than the physical gap into which they were moved. We therefore assigned the first SNP of the block to a position 10 kb downstream from the previous SNP on the correct linkage group. We then recalculated genomic coordinates for the rest of the chromosome based on the marker distances within the translocated segment. The last SNP of the block was also given a 10 kb cushion between itself and the next SNP on the correct linkage group.

Only a portion of the SNPs on the array had genetically mapped positions. 'Unmapped' SNPs (those with a physical position but no genetic position) bounded by two mapped markers were moved along with their mapped neighbors if both anchoring mapped markers were also moved. However, it is unclear whether unmapped SNPs just outside of these anchors should be kept in place or moved along with the adjacent SNPs. Since most inversions were small relative to the genetic map (and would therefore still fall in the same window of a sliding window analysis), these SNPs were left in place. However, markers bordering translocations were removed to ensure there were no markers mapped to the incorrect linkage group. Unmapped SNPs were then assigned a genetic position by linear interpolation of genetic vs. physical distance using the *approx()* function in R ([www.R-project.org](www.R-project.org)). The IBM genetic map distances were then converted to single-meiosis map distances using the formulae of Winkler et al. (Winkler et al., 2003). Finally, SNPs located at physical positions outside of those bounded by the genetic map (such as the telomeres) were assigned the genetic position of their nearest mapped neighbor. Since moved segments were arbitrarily joined 10 kb from their nearest genetic neighbor, we acknowledge that the physical positions of these markers are only estimates. However, the estimated junctions are small relative to the genetic windows used for our analysis. The final map used is provided as supplemental material (Supplementary File S1)

**Haplotype phasing:** Although the genotypes of the plants from each population are unphased, the homozygous genotypes of the founders and derived inbreds provide excellent prior information for a probabilistic estimation of genotype phase in the populations. We therefore used fastPHASE (Scheet and Stephens, 2006) to estimate the genotype phase of each plant. To estimate the error in phasing, we created test cases by combining the genotypes of two derived inbreds into a hypothetical "F1 hybrid" of unknown phase. This F1 was presented to fastPHASE with the rest of the data, except that its parent inbreds were removed. Analyses of several hypothetical F1's from different cycles of selection revealed very low phasing error rates (Table S3). Therefore the phased genotypes of cycle 0 plants were used as the starting data for simulations (see below).

**Diversity and principal component analysis:** Heterozygosity (H) was measured as:

$$H = 2p(1-p)$$

where $p$ and $(1-p)$ are the frequencies of the two SNP alleles. $F_{ST}$ (Hudson et al., 1992) was calculated using the *HBKpermute* program in the analysis package of the software library libsequence (Thornton, 2003). All results were plotted using the R package *ggplot2* (Wickham, 2009). We conducted PCA by singular value decomposition, as described in (McVean, 2009).

**Simulations:** Our simulation sought to model the effects of genetic drift in the Iowa RRS experiment independent of any selection, and our model thus closely followed the published methods of the Iowa RRS (Penny and Eberhart, 1971; Keeratinijakal and Lamkey, 1993). Starting individuals in each population were constructed by randomly sampling two distinct haplotypes with replacement from the phased haplotypes of cycle 0. In the actual random mating scheme used in the Iowa RRS experiment, a single pairing could only contribute four gametes to the next generation (two kernels each from two ears), and our simulation reflects this. Advanced cycles were simulated by randomly mating gametes from self-fertilized plants of the previous cycle until 10 new individuals were created. The first cycle involved two rounds of random mating, whereas all subsequent cycles used one round. After cycle 5, the process employed two rounds of selfing instead of one. After cycle 7, the population size was increased from 10 to 20. At cycles 4, 8, 12, and 16, the plants were randomly mated to match the sample size of the observed data. The genotypes of these simulated random matings are the final results of each simulation that were analyzed in the same way as the observed data.

All recombination events in the simulations were carried out in R with the *hypred* software package (cran.r-project.org/web/packages/hypred/). The simulations were executed in parallel on a computing cluster, with unique random number seeds drawn for each simulation. Statistics were calculated for each simulation using the same formula as with the experimental data. We used non-overlapping sliding windows of equal genetic distance to account for the non-independence of markers in low-recombination regions when calculating measures of significance. For the haplotype-based, single-locus simulations, recombination was simply replaced with binomial sampling of two alleles.

RESULTS

**Increase in population structure and loss of genetic diversity:** Founder inbreds and samples from cycles 0, 4, 8, 12, and 16 were genotyped at 39,258 SNPs that passed a set of quality filters and could be assigned collinear genetic and physical map positions (See Materials and Methods for details). Change in population structure throughout the Iowa RRS experiment can be observed visually by a principal component analysis (PCA). A joint analysis of individuals from all the selection cycles (Figure 1) clearly separates the BSSS and BSCB1 populations along the first axis of variation, with increasing separation as the experiment progressed. The second axis of variation primarily separates the cycles from one another within each population. There is no separation between the founders of the two populations, but at cycle 0, the BSSS population shows more divergence from the founders than does BSCB1. This could be due to drift during either the population's construction or its subsequent maintenance. Structure continued to develop within each population over the course of the experiment. There is an especially wide gap between cycles 4 and 8, which correlates with the addition of an extra generation of self-pollination prior to selection at each cycle. The distance between cycles then decreases dramatically after cycle 8, and this corresponds to doubling in effective population size.

No new genetic material was intentionally introduced into either population after the experiments' inception, so the substantial increase in genetic distance could only arise from the loss of genetic diversity within each population. Consistent with previous studies of the Iowa RRS, genome-wide genetic diversity decreased steadily across cycles of selection in both populations (Figure 2). The decrease in heterozygosity (H) was much smaller when the two populations are combined, indicating the loss of different alleles within BSCB1 and BSSS. This leads to strong genetic differentiation, as $F_{ST}$ rose by a factor of ten between the founder lines and the populations at cycle 16.

We noticed an irregular increase in the number of polymorphic markers between BSSS cycles 4 and 8 (Table S2). All of these newly polymorphic markers were present at extremely low frequency and were spread among various individuals. This may represent a series of minor alleles that were not captured in our sample of cycle 4 individuals and thus appeared to 'resurface' at cycle 8. Alternatively, the pattern may be the result of minor contamination at some point in the population's history. It was observed that an allele of the *sugary* gene associated with sweet corn appeared in the population at this time (O. S. Smith, personal communication), suggesting contamination may be the cause. However, the low frequency of the 'new' alleles (typically only one or two alleles out of 72 possible in 36 diploid samples) means their effect on the population diversity is minimal. We did not attempt to incorporate this contamination into our simulation approaches, as it only makes our tests for low heterozygosity slightly more conservative.

**Fixation of large genomic regions:** Figure 3 shows heterozygosity varying along the genome across cycle 16 of each population. Of particular note in this figure are extremely large regions of zero or near-zero heterozygosity spanning tens of megabases that lie in the pericentromeric regions. These regions experience low rates of meiotic recombination, which creates an expanded physical map relative to their genetic length (Ganal et al., 2011). In general, the majority of fixed haplotype segments are small in genetic space (most are less than 2 cM) regardless of their physical size. One exception is a region of fixation that spans 7 cM on chromosome 1 in the BSCB1 population.

The sheer physical size of the pericentromeric regions yields extremely high marker density on the genetic map, allowing for clear resolution of haplotype phasing and recombination breakpoints. To further examine the fixation in these regions, we computationally imputed haplotype phase in the BSSS and BSCB1 populations and used the phased data to track haplotype frequencies and founder of origin. In most cases, these fixed haplotypes can be traced back to single founder inbreds. For example, in BSSS a 60 Mb, 1.2 cM region of chromosome 9 became fixed by cycle 12 (Figure 4) and traces back to the founder Os420. In BSCB1, a 60 Mb, 0.7 cM region became fixed on chromosome 4 and traces back to A340 (Figure 5). Table 1 gives a summary of the large genomic regions that have become fixed or nearly fixed by cycle 16. These regions represent large blocks of linked alleles that show no evidence of recombination since at least the development of the founding inbred lines in the 1920's and 1930's.

**The role of genetic drift:** There is clear evidence of phenotypic improvement in response to selection in the Iowa RRS populations (Smith, 1983; Keeratinijakal and Lamkey, 1993; Schnicker and Lamkey, 1993; Holthaus and Lamkey, 1995; Brekke et al., 2011a; Brekke et al., 2011b; Edwards, 2011) and large changes in genetic structure indicated by molecular markers. A central issue for these maize populations and others like them is whether the changes observed at the molecular level are caused directly by selection on phenotype, or indirectly due to the genetic drift that selection imposes through inbreeding and small effective population sizes. To gauge the roles of selection and drift, we conducted simulations of the crossing and selection schemes used in the RRS experiment, with genetic recombination modeled on a framework map of a cross between B73 (a BSSS derived line used to construct the reference genome) with Mo17 (an inbred with some co-ancestry to both the BSSS and BSCB1 founder germplasm)(Gerdes et al., 1993; Lee et al., 2002). Selections were executed at random in each simulation, so the patterns observed across simulations represent the expected distribution of effects caused by recombination and genetic drift. We conducted 10,000 simulations, sampled individuals from each cycle equal in number to the observed samples, measured heterozygosity, and compared the results to the observed data.

Averaged across the genome, the vast majority of the reduction in diversity observed in both populations can be attributed to genetic drift. However we do observe deviations from the simulated values (Figure S1). The observed data show higher than expected heterozygosity at cycle 4, suggesting selection for heterozygosity in early cycles, a deviation from simulated and actual breeding practices, or an under-sampling of the diversity present in the original cycle 0 populations. Nevertheless, as the cycles progress heterozygosity falls more than expected in both populations and the observed values at cycle 16 are significantly lower than the simulated data. Simulations across a number of different marker densities were consistent with this result (data not shown).

To examine the behavior of specific regions, we also calculated the observed and simulated results for each 2 cM segment of the genome. The dynamics observed across most of the genome are largely insensitive to changes in window size (we tested from 2-4 cM, data not shown), and are consistent with strong genetic drift imposed by selection practices. A subset of loci were flagged as significant, and these loci almost always overlapped regions of fixation or near-zero heterozygosity in one population. However, in these regions the values obtained by simulation are often quite low, such that the extreme simulated values sometimes lie only a few percentage points away from the observed value (Supplementary File S2), indicating that drift alone can explain most of the drop in diversity.

Since the population size of the Iowa RRS is small (10-20), many biallelic SNPs should fix by chance regardless of their starting minor allele frequencies. The deviation from simulation arises not from changes in allele frequencies per se, but rather from the fixation of linked markers across larger than expected genetic distances. The validity of significance cutoffs therefore depend on the accuracy of our genetic map, and the maize genetic map is known to vary among genetic backgrounds across short genetic distances (< 5cM) (McMullen et al., 2009). Deviations of the observed and simulated results could be due to selection, variation / inaccuracy in the genetic map, or a combination of these factors.

To explore the roles of selection and drift independent of the genetic map, we returned to the large regions of fixation in the centromeres which showed no recombination across the full RRS experiment. Given the lack of recombination, each of these regions can be analyzed by the simulation of a single locus, and the high density of markers allows the clear resolution of the individual haplotypes. We used the computationally phased data to measure the frequency of the fixed haplotype at each cycle, and assessed the probability of observing the fixation event given the initial frequency. The BSSS chromosome 9 haplotype fixed at cycle 16 was at low frequency at cycle 0 (7 out of 68 haplotypes), but increased rapidly in frequency by cycle 8 (66 out of 70 haplotypes). Simulation of the haplotype as a single locus in the RRS experiment produces this increase in frequency in only 3.9% of 1000 independent simulations, whereas the haplotype was lost in over 80% of the simulations. In BSCB1, a 30 Mb, 1.6 cM region of chromosome 2 became nearly fixed by cycle 8 (67/70) despite a prevalence of 4/72 at cycle 0, which occurred 1.5% of the time by simulation. Although these results suggest that selection may have pushed these haplotypes to fixation, the fact that fixation of such a rare haplotype still occurred in some simulations speaks to the strong genetic drift imposed upon the BSSS and BSCB1 populations. Interestingly, each of these two genomic regions harbored a different cycle 0 haplotype at higher frequency, but these higher-frequency haplotypes were subsequently lost within the RRS population. In other cases, the haplotypes which eventually fixed were at moderate frequency in the cycle 0 populations and drift to fixation in the majority of simulations.

DISCUSSION

Reciprocal recurrent selection serves as a model for the method of hybrid maize improvement employed broadly in North America (Duvick et al., 2004). In the Iowa RRS experiment, the BSSS and BSCB1 populations steadily lost genetic diversity as they become more diverged from one another. Principal component analysis shows that as the effective population size and the rates of inbreeding were altered, the rates of change in population structure were altered as well. The population structure and divergence between BSSS and BSCB1 can be largely reproduced by simulation without any selection. The trends in these populations support the hypothesis that the majority of the genetic structure observed can be attributed to genetic drift alone, despite effective selection for phenotypic improvement. This drift was driven by inbreeding through self-pollination and the small effective population sizes used to select a limited number of high-performing, potentially related individuals at each cycle. The trends observed in the Iowa RRS are also a common theme in similar recurrent selection experiments (Romay et al., 2012; C.

Lamkey and A. Lorenz, personal communication). Increased population structure due to increased inbreeding and lowered effective population size are visible in the modern 'stiff-stalk' and 'non stiff-stalk' heterotic groups as well (van Heerwaarden et al., 2012). Given the lack of strong selection signals in each case, genetic drift has most likely played a role in the current genetic structure of modern maize.

Several key inbreds in the 'stiff-stalk' heterotic group, B73, B37 and B14, were derived from the BSSS population (Darrah and Zuber, 1986; Troyer, 1999). B37 and B14 were derived from cycle 0, and B73 was derived from a half-sib recurrent selection program also started with the BSSS population. We examined these three inbreds at the pericentromeric regions listed in Table 1, and found that in most cases they carry different haplotypes from those that rose to high frequency in the RRS experiment. Near isogenic lines that substitute these centromeric haplotypes from key inbreds could be used to determine if the haplotypes at high frequency in cycle 16 of BSSS provide a phenotypic advantage when crossed with BSCB1.

Both our study and previous work (Labate et al., 1997) identified genetic differentiation between cycle 0 of BSSS and the founder inbreds. The loss of diversity in the cycle 0 population likely led to the loss of some rare alleles and haplotypes. BSSS was maintained for several years before RRS was initiated, and the differentiation could have occurred due to low effective population size during maintenance. If so, then this drift will have impacted the trajectory of the RRS program and of the key inbreds derived from BSSS. However, cycle 0 has also undergone maintenance since the beginning of the RRS experiment, so some genetic drift may have occurred during this time as well.

Although drift can explain most of the genetic structure genome-wide, phenotypic data provide clear evidence that selection has altered the frequencies of favorable alleles in the BSSS and BSCB1 populations. Numerous experiments have shown that the selected populations and the hybrids formed from them are superior to their cycle 0 ancestors (Smith, 1983; Keeratinijakal and Lamkey, 1993; Schnicker and Lamkey, 1993; Holthaus and Lamkey, 1995). Improvement occurred not only for hybrid yield, but concurrently for plant architecture and tolerance to high-density planting in both hybrids and inbreds (Brekke et al., 2011a; Brekke et al., 2011b; Edwards, 2011; Lauer et al., 2012). Since these same trends are observed across wider North American germplasm (Tollenaar, 1989; Duvick, 2005), identification and characterization of the loci conferring phenotypic improvement in BSSS and BSCB1 could reveal some of the genetic mechanisms by which maize yield has steadily improved over the decades. We find that the genomic regions of extremely low diversity evident at cycle 16 are unlikely to be produced by simulation, and heterozygosity falls more than expected across the genome as a whole. However, simulated and observed values are often close, so overall the signature of selection is difficult to detect at any given locus. This is due in large part to a limitation of statistical power imposed by inbreeding and low effective population sizes, though uncertainty about the fine-scale recombination map also plays a role. We show that an identity-by-descent, haplotype based approach provides additional power as it can distinguish between the fixation of rare and common haplotypes. However, the results of single locus simulations are sensitive to sampling error and drift caused by population maintenance. In this case, it is difficult to assess significance with only a single population to test. This type of analysis would be more effective across several replicated populations, which can control for genetic drift due to the independence between the selections in each replicate (C. Lamkey and A. Lorenz, personal communication).

van Heerwaarden et al., 2012 assessed variation and tested for selection across a wider range of North American maize germplasm using the same SNP array. Although selected loci were identified in that study, the overall effects on genome diversity and haplotype patterns were relatively small compared with the impact of domestication. That study was a wide survey across many germplasm sources. The limited impact of selection could arise because different haplotypes and loci are selected in different breeding programs. In support of this idea, we find that the most likely targets of selection between the BSSS and BSCB1 populations occur at non-overlapping loci. This non-overlap between putatively selected loci in complementary populations has also been observed in commercial breeding programs (Feng et al., 2006).

The observation that the same targets of selection are not observed in the opposing heterotic populations bears implications for the genetic mechanisms responsible for heterosis and the success of maize hybrids. Classic overdominance models of heterosis predict that at a single locus, two distinct alleles confer heterozygote advantage when combined, while the dominance model predicts that heterosis is driven by dominance effects and the complementation of linked alleles in low-recombination regions (dominance or pseudo-overdominance). In the case of true overdominance, we expect selection should lead to decreased heterozygosity at a locus in both populations as complementary haplotypes are fixed in each group. We find no evidence of this genetic phenomenon. The observed pattern instead favors a dominance model, where fixation of a haplotype in one population simply selects against that same haplotype in the other population. Because most deleterious alleles are rare in both heterotic groups (S. Mezmouk and J. Ross-

Ibarra, unpublished results), most haplotypes in the second population will have a different suite of deleterious variants and will complement the fixed haplotype reasonably well such that selection will have little impact on diversity in the second population. Our data are consistent with such a process having been important for hybrid improvement in the Iowa RSS experiment, and the lack of extreme differentiation seen over time across US Corn Belt germplasm is consistent with a role for such selection in maize hybrid improvement in general (van Heerwaarden et al., 2012).

Our results are consistent with the dominance and pseudo-overdominance models, but we caution that the exact outcome of particular models will depend strongly on the effects of selection, drift, and the frequencies of beneficial alleles. Disentangling these factors remains a major challenge that will require careful simulation of maize breeding using a wider range of parameters. This is especially true because in a model of hybrid complementation, genetic drift in one population can alter the selective value of alleles in the other population. When drift and selection interact in such a manner, tests for selection that attempt to independently partition the effects of each force may not provide the full picture. Furthermore, uncertainties in the fine-scale relationship between genetic and physical distance make it difficult to assign significance in forward-simulation approaches. In the end, the best test for selection of specific genomic regions will ultimately be conducted by phenotypic observation in the field in balanced tests of different haplotypes. In the case of individual breeding programs, genome-wide genotyping such as we conducted here can identify the lines carrying the recombinant haplotypes and introgressions necessary to conduct such an experiment.

**Table 1.** Ancestry of haplotypes fixed in the cycle 16 population.

| Population | Chromosome | Genetic Interval (cM) | Physical Interval (Mb) | Founder Inbred | Derived Lines with Haplotype |
|---|---|---|---|---|---|
| BSSS | 3 | 40.2-41.4 | 67.7-123 | CI187-2 | B94 |
| BSSS | 3 | 43-48.5 | 129.2-157.1 | ND[a] | B89, B94 |
| BSSS | 4 | 39.7-42.1 | 39.9-82.7 | CI187-2 | B89, B94, B67, B72, B39, B43 |
| BSSS | 9 | 30.6-33.5 | 20.8-26.6 | Oh7[b] | B89, B94, B43, B17, B72, B84, B67 |
| BSSS | 9 | 34.3-35.6[c] | 30.8-90.4[c] | Os420 | B89, B94 |
| BSCB1 | 2 | 50.2-51.8 | 80.6-114.5 | CC5 | B90, B91, B95, B97, B99 |
| BSCB1 | 4 | 42.1-42.8 | 82.7-140 | ND[a] | B90, B95, B97 |
| BSCB1 | 8 | 46.1-50.7 | 125.1-145.6 | P8 | B90, B97, B91, B99, B54 |

[a] ND = Not Determined (Either a recombinant haplotype or originates from an un-genotyped founder)

[b] Although Oh7 is a BSCB1 founder, it is a descendant of CI.540, an un-genotyped BSSS founder. BSSS segments matching Oh7 presumably derive from CI.540

[c] Founders Ind_B2 (BSSS), CI187-2 (BSSS), R4 (BSCB1), and I205 (BSCB1) are all IBD at this region of chromosome 9

ACKNOWLEDGEMENTS

We thank O. S. Smith and members of the Ross-Ibarra lab for comments on earlier versions of the manuscript. J.P.G received support for this research as a Merck Fellow of the Life Sciences Research Foundation. This research was supported by the National Science Foundation (IOS-0820619) and funds provided to USDA-ARS (MDM). Names of products are necessary to report factually on available data: however, neither the USDA, nor any other participating institution guarantees or warrants the standard of the product and the use of the name does not imply approval of the product to the exclusion of others that may also be suitable.

FIGURE LEGENDS

**Figure 1.** Principal component analysis of the SNP data from Iowa RRS. The axes represent the first two eigenvectors from an analysis of cycles 0-16, with projection of the founder lines onto the vector space. The variation explained by each eigenvector is given in parentheses on the axes. The populations steadily diverge at increasing cycles, with less distinction visible between the founder groups. The comparatively large distance between cycles 4 and 8 corresponds to a switch from one to two generations of selfing at each cycle. The smaller separation between cycles 8-16 corresponds to an increase in effective population size from ten to twenty. The BSSS cycle 0 population has drifted away from the BSSS founders, despite the absence of intentional selection during the creation and maintenance of cycle 0.

**Figure 2. Heterozygosity and $F_{ST}$ across cycles**. $H$ (left panel) and $F_{ST}$ (right panel) plotted as a function of selection cycle in each population.

**Figure 3. Heterozygosity ($H$) at cycle16 across all ten chromosomes in each population.** $H$ is calculated on 15-marker sliding windows with 5 marker steps. Each point is plotted at the midpoint of the 15-marker window.

**Figure 4. Heterozygosity ($H$) across chromosome 9 in each cycle of BSSS plotted on the physical (A) or genetic (B) map.** H is calculated on 15-marker sliding windows, with 5 marker steps between each calculation. Each data point is color-coded based on a linear transformation of recombination rate (red = low recombination rate). Shaded regions represent 2 cM windows

found to have heterozygosity values significantly lower than expected by simulation at a given cycle. Light shading denotes significance at $P < 0.0025$, whereas the darker shading represents significance at $P < 0.001$. Comparing the physical and genetic maps reveals that the regions of fixation are extremely small genetically, but some contain much of the physical content of the chromosome.

**Figure 5. Heterozygosity (H) across chromosome 4 in each cycle of BSCB1 plotted on the physical (A) or genetic (B) map.** $H$ is calculated on 15 marker-sliding windows, with 5 marker steps between each calculation. Each data point is color-coded based on a linear transformation of recombination rate (red = low recombination rate). Shaded regions represent 2 cM windows found to have heterozygosity values significantly lower than expected by simulation at a given cycle. Light shading denotes significance at $P < 0.0025$, whereas the darker shading represents significance at $P < 0.001$.

**Figure S1. Heterozygosity in each population, observed vs. simulated data.** Heterozygosity was calculated as the average across all markers genome-wide in the BSSS (A) and BSCB1 (B) populations. The observed data is marked by the red line, the simulations by gray dots, and the median of the simulations by a green dot.

**Figure S2. Heterozygosity (H) at cycle16 across all ten chromosomes in each population, calculated on 15-marker sliding windows with 5 marker steps as in Figure 3.** Heterozygosity values in BSSS (blue dots) and BSCB1 (red dots) are superimposed in one panel. 2 cM windows

of low heterozygosity that were reproduced in less than 10 of 10,000 simulations (P<0.001) are shaded in light blue (BSSS) or pink (BSCB). At this cutoff, there is no overlap between highlighted windows in the two populations. (A) Physical map. (B) Genetic map.

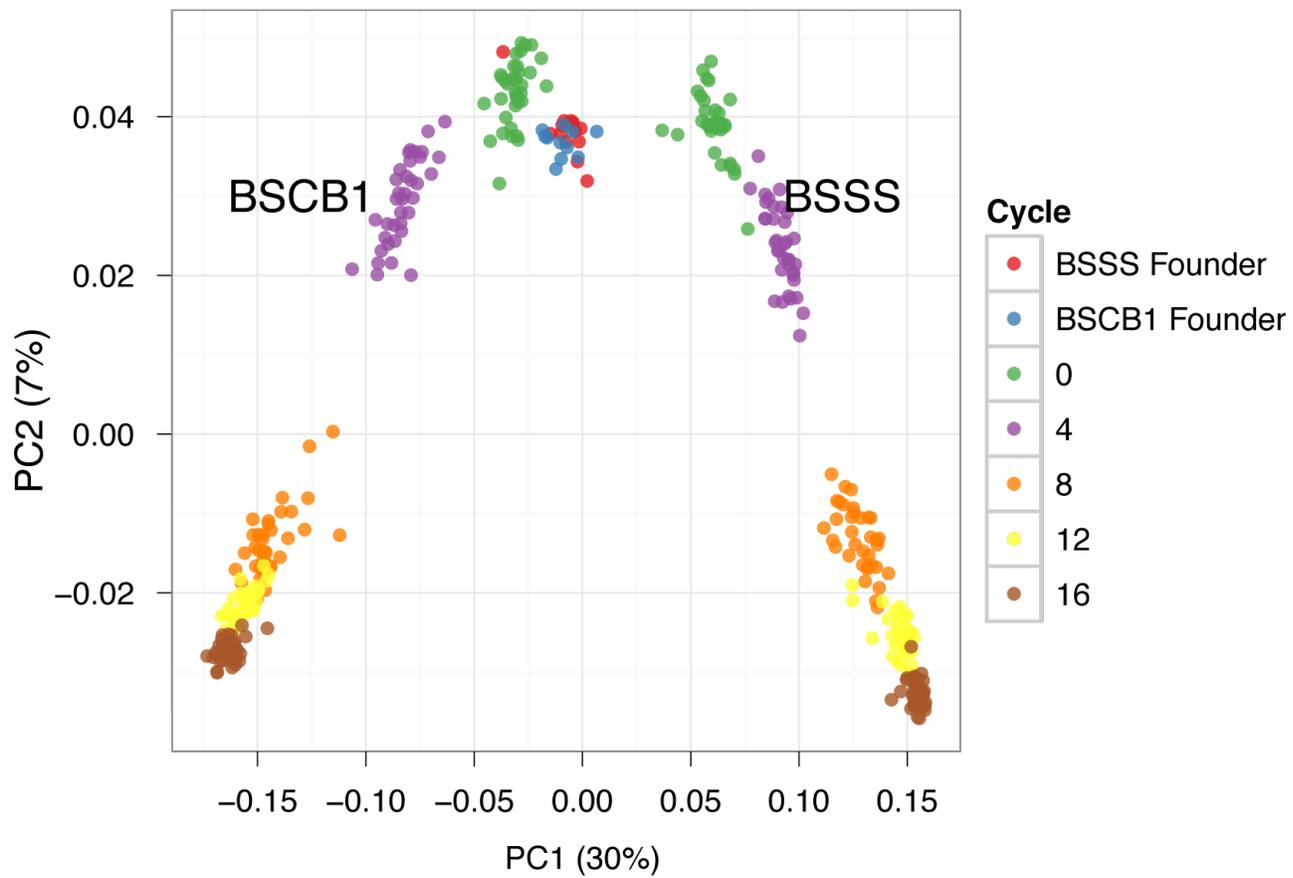

Figure 1

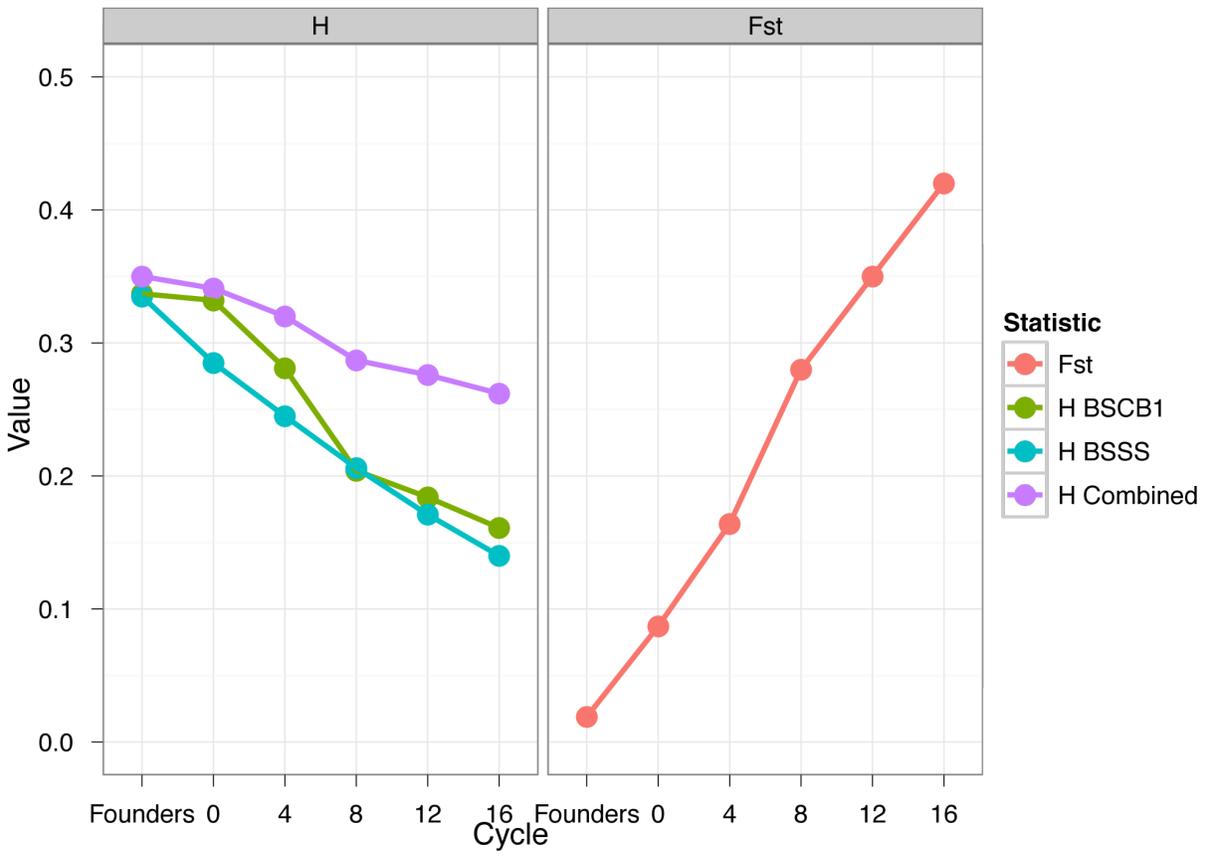

Figure 2

# Figure 3

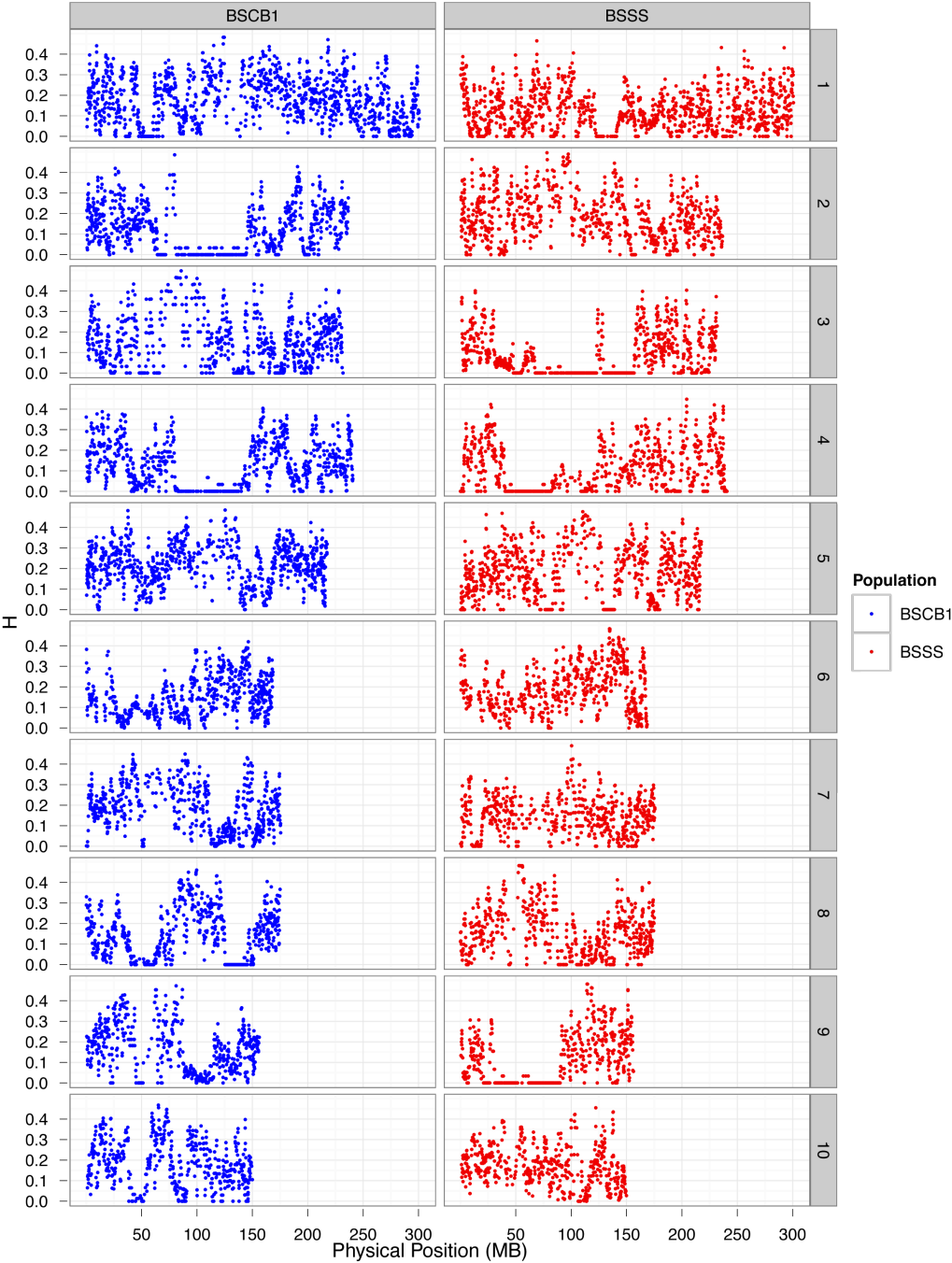

# Figure 4

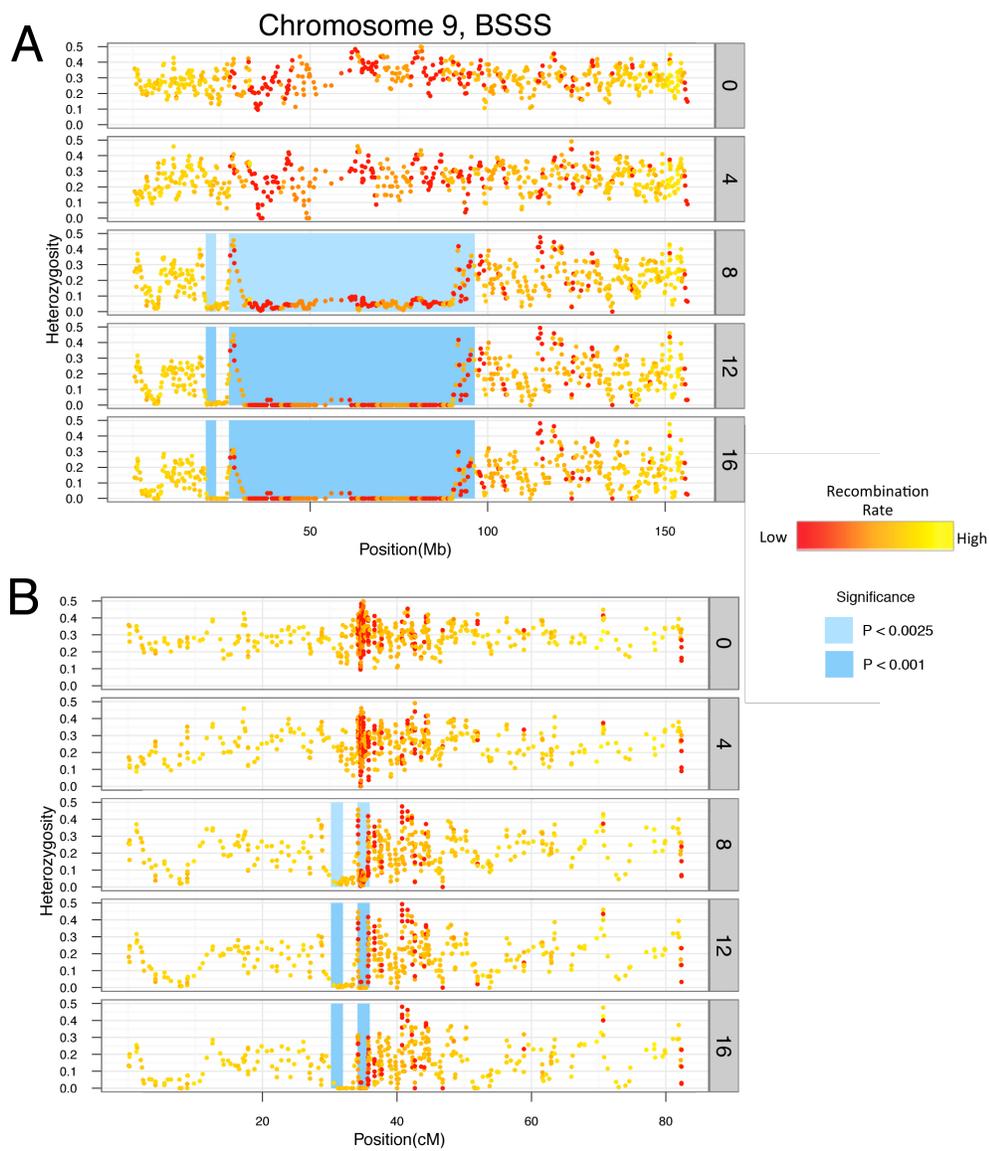

# Figure 5

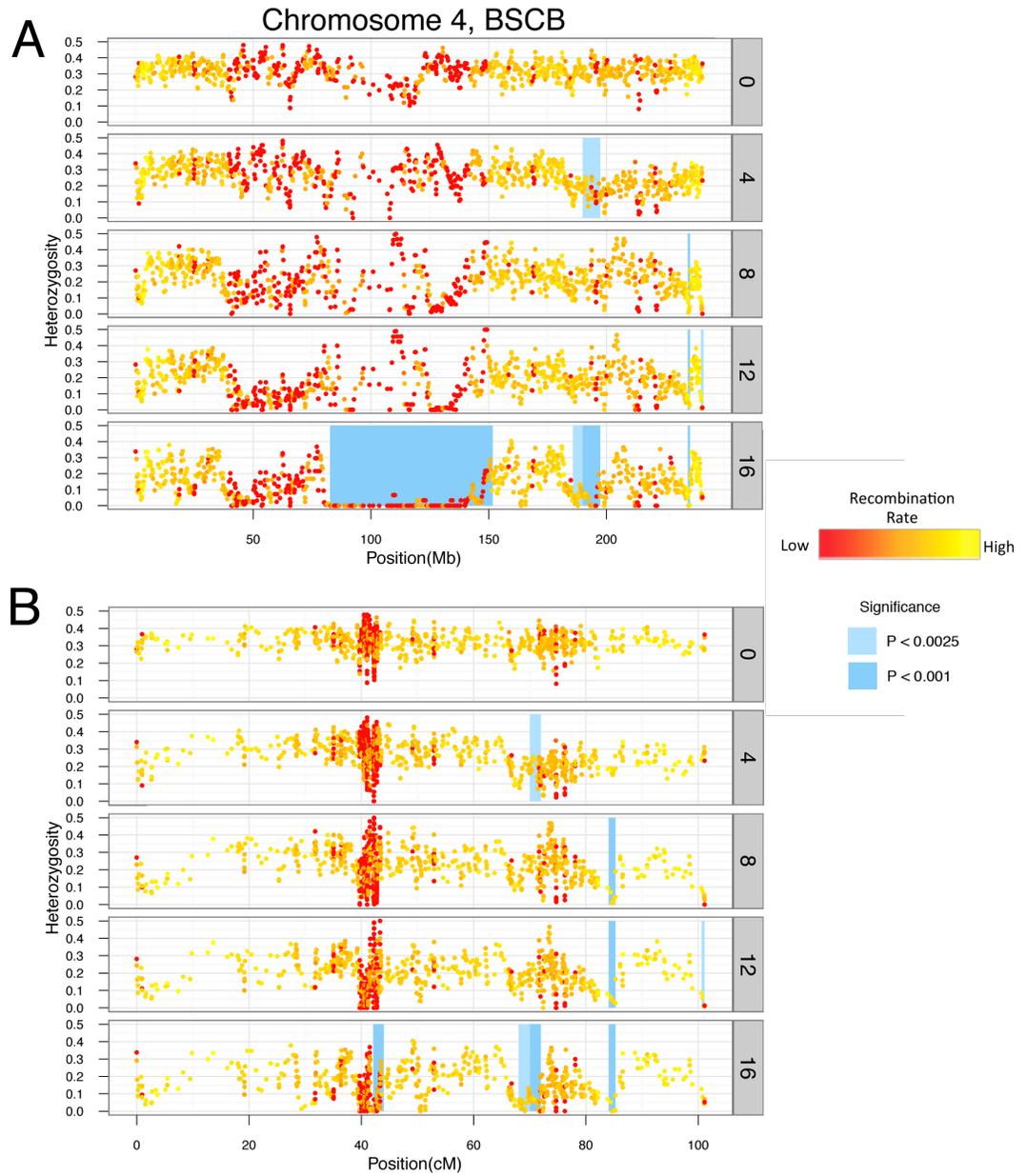

# Figure S1

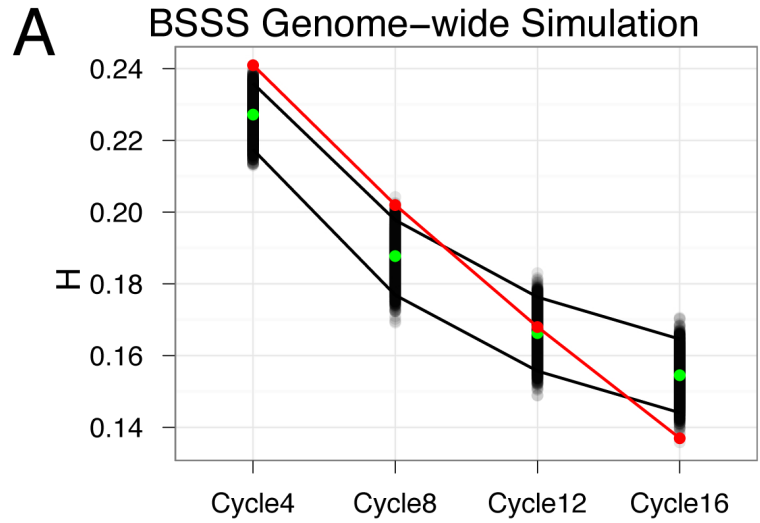

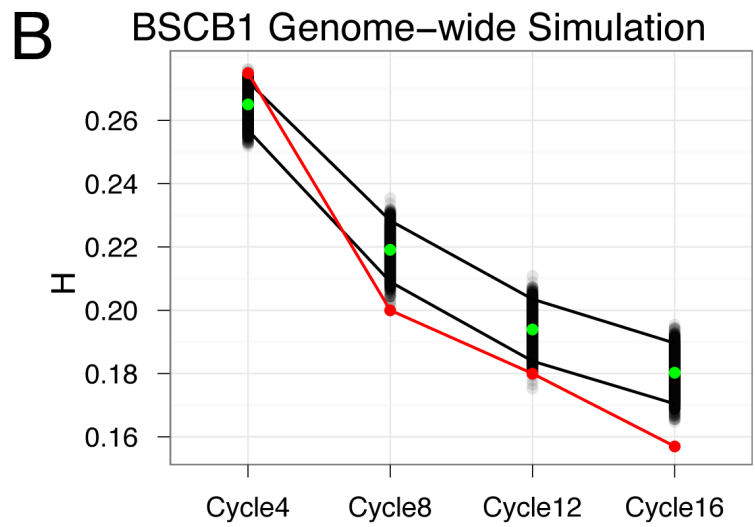

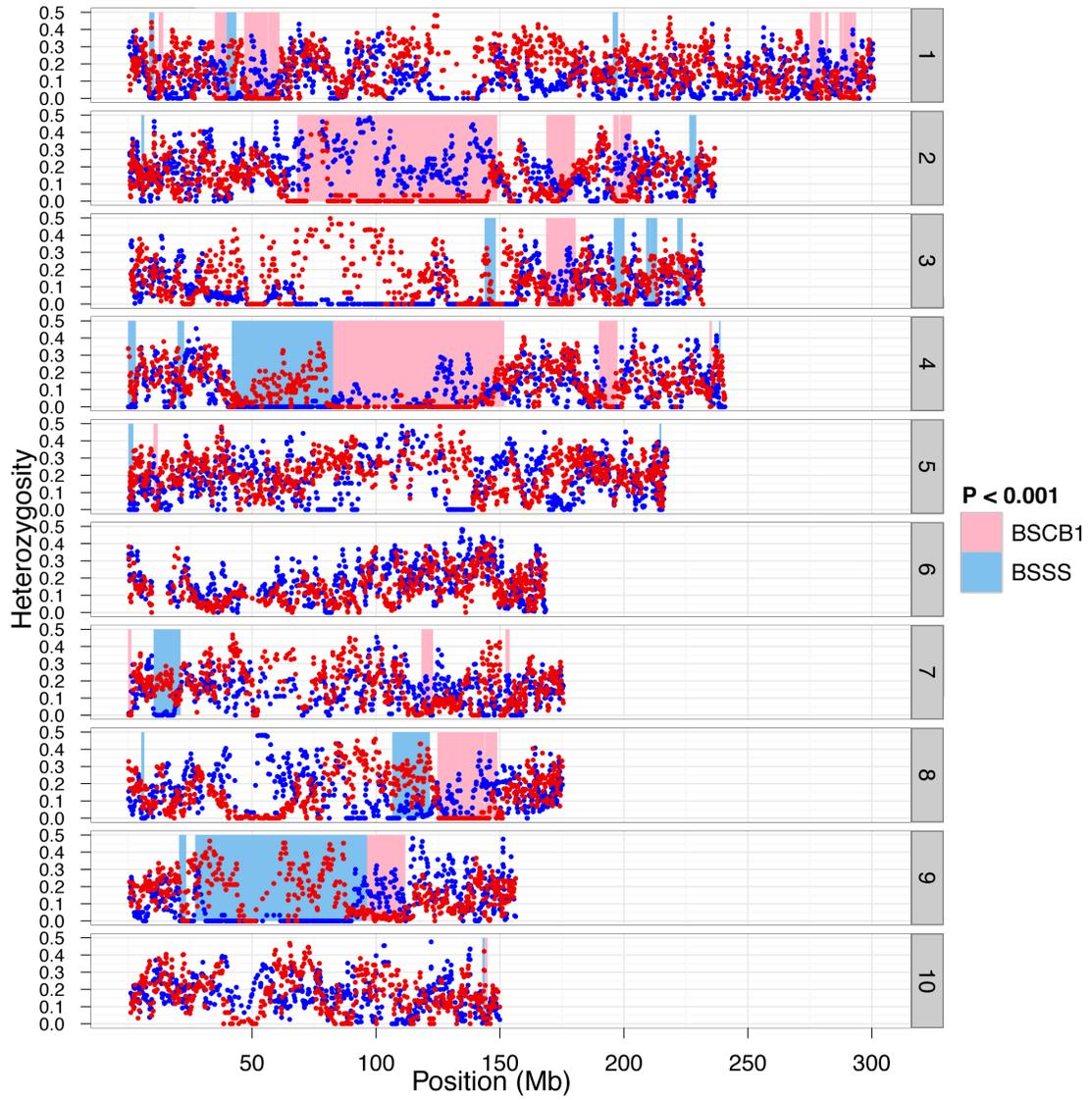

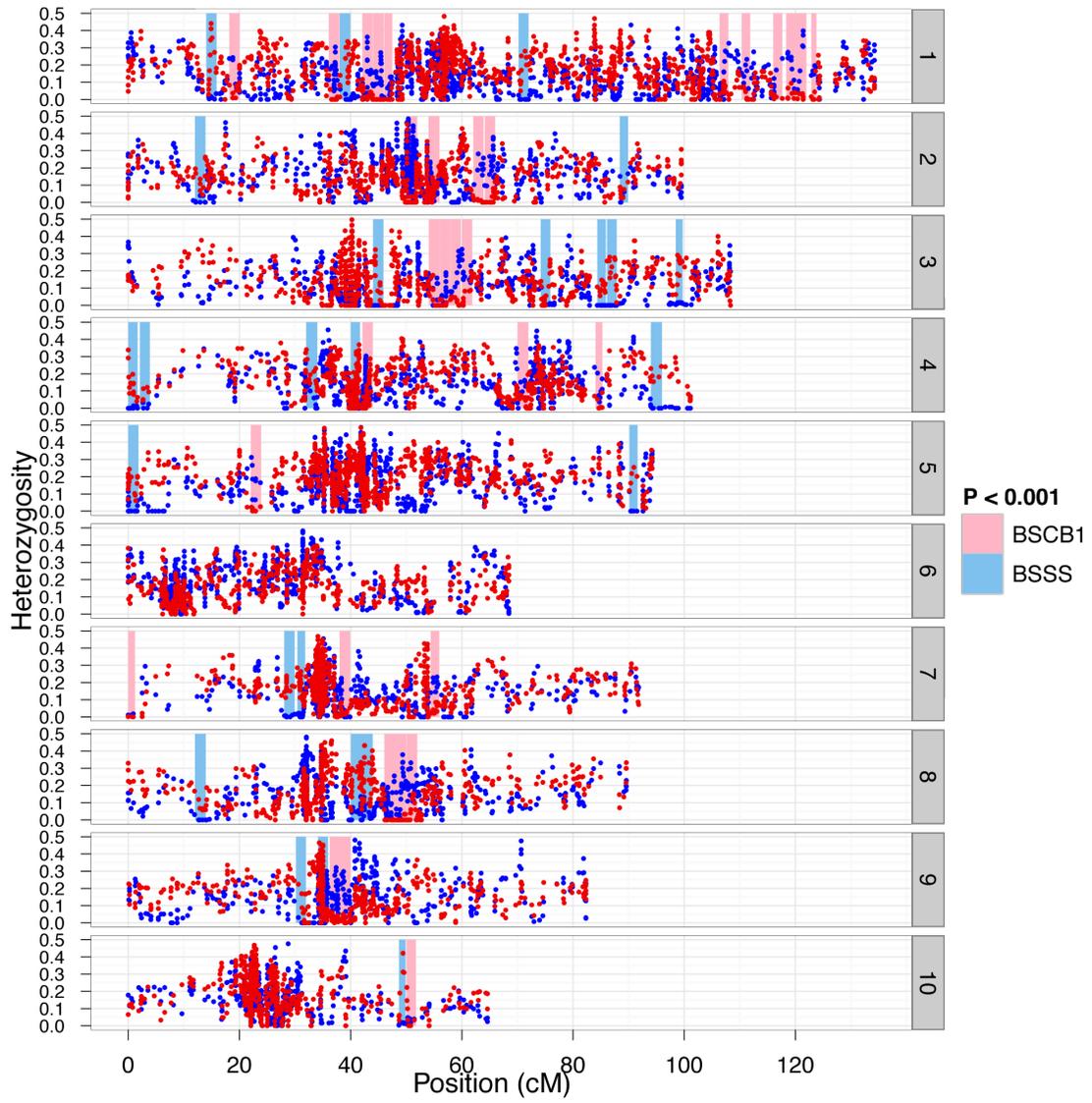

Figure S2 B

**Table S1.** Plants and Lines Genotyped in this study.
Backgrounds of the founder inbreds are from Hagdorn et al. 2003.

### Plants from the selection cycles

| Population | Cycle | # plants |
|---|---|---|
| BSSS | 0 | 34 |
| BSSS | 4 | 36 |
| BSSS | 8 | 35 |
| BSSS | 12 | 36 |
| BSSS | 16 | 36 |
| BSCB1 | 0 | 36 |
| BSCB1 | 4 | 36 |
| BSCB1 | 8 | 35 |
| BSCB1 | 12 | 36 |
| BSCB1 | 16 | 36 |

### BSSS Founders Genotyped

| Inbred | Background / Pedigree |
|---|---|
| Ind_Tr9_1_1_6 | Reid Early Dent (Troyer Strain) |
| Oh3167B | Echelberger Clarage |
| I224 | Iodent |
| Ind467(744) | Reid Medium |
| CI.187-2 | Krug-Nebraska Reid x IA Gold Mine |
| Os420 | Osterland Yellow Dent |
| I159 | Iodent |
| A3G-3-1-3 | BL345BxIAI129 |
| Ind_Fe2_1073* | Troyer Reid (Early) |
| Ill_Hy | IL High Yield |
| Ill_12E | unknown |
| Ind_AH83 | Funk 176A |
| Ind_B2* | Troyer Reid (Late Butler), parent of a founder |
| LE23 | IL Low Ear |

*Parents of unavailable line F1B1

### BSCB1 Founders Genotyped

| Inbred | Background / Pedigree |
|---|---|
| I205 | Iodent |
| Oh51A | [(OH56xWf9)Oh56] (Wooster Clarage x ?) |
| A340 | 4-29 x 64 (Silver King x Northwestern Dent) |
| Ill_Hy | IL High Yield |
| Oh33 | Clarage |
| Oh07 | C.I.540xIll.L |
| R4 | Funk Yellow Dent |
| Oh40B | eight line LSC composite |

| | | |
|---|---|---|
| P8 | | Palin Reid |
| L317 | | LSC |
| CC5 | | Golden Glow (W23) |

### Founders not analyzed

| Inbred | Group | Reason |
|---|---|---|
| CI.540 | BSSS | heterozygous genotype |
| F1B1 | BSSS | unavailable |
| CI.617 | BSSS | unavailable |
| WD456 | BSSS | unavailable |
| K230 | BSCB1 | source segregates phenotypically |

### Derived Lines Used

| Inbred | Group | Cycle | Notes |
|---|---|---|---|
| B10 | BSSS | 0 | thrown out; poor data quality |
| B42 | BSCB1 | 0 | |
| B14A | BSSS | 0 | Cuzco x B14 |
| B43 | BSSS | 0 | |
| B10 | BSSS | 0 | |
| B37 | BSSS | 0 | |
| B44 | BSSS | 0 | |
| B17 | BSSS | 0 | |
| B69 | BSSS | 0 | |
| B39 | BSSS | 0 | |
| B90 | BSCB1 | 7 | |
| B40 | BSSS | 0 | |
| B54 | BSCB1 | 0 | |
| B78 | BSSS | 8 | from half-sib recurrent selection program |
| B72 | BSSS | 3 | from half-sib recurrent selection program |
| B84 | BSSS | 7 | from half-sib recurrent selection program |
| B94 | BSSS | 8 | |
| B99 | BSCB1 | 10 | |
| B11 | BSSS | 0 | |
| B89 | BSSS | 7 | |
| B95 | BSCB1 | 7 | |
| B91 | BSCB1 | 8 | |
| B67 | BSSS | 0 | |
| B73 | BSSS | 5 | from half-sib recurrent selection program |
| B97 | BSCB1 | 9 | |

**Table S2.** Evidence of minor contamination between cycles 4 and 8 in BSSS.

| SNPS polymorphic in cycle: | But not cycle: | in BSSS | in BSCB1 |
|:---:|:---:|:---:|:---:|
| 4 | 0 | 290 | 225 |
| 8 | 0 | 2242 | 319 |
| 12 | 0 | 1227 | 181 |
| 16 | 0 | 344 | 128 |
| 8 | 4 | 4499 | 1081 |
| 12 | 4 | 2328 | 669 |
| 16 | 4 | 551 | 477 |
| 12 | 8 | 1821 | 1405 |
| 16 | 8 | 178 | 997 |
| 16 | 12 | 542 | 666 |
| 0 | Founders | 1202 | 1707 |
| 4 | Founders | 822 | 885 |
| 8 | Founders | 1201 | 798 |
| 12 | Founders | 816 | 550 |
| 16 | Founders | 445 | 460 |

Heterozygosity ($H$) of SNPS present in cycle 8 but not cycle 4

| Statistic | BSSS | BSCB |
|:---:|:---:|:---:|
| Mean | 0.052 | 0.052 |
| Median | 0.028 | 0.028 |

**Table S3.** Switch error rates from computationally phasing 'hybrids' simulated from derived inbred lines.

| Simulated 'Hybrid' | Derived Lines Used as Priors | Population | Possible Switches | Switch errors | Rate |
|---|---|---|---|---|---|
| B11xB67 | all | BSSS | 12195 | 41 | 0.003 |
| B17xB44 | all | BSSS | 12230 | 58 | 0.005 |
| B39xB37 | all | BSSS | 11485 | 129 | 0.011 |
| B43xB69 | all | BSSS | 12266 | 26 | 0.002 |
| B73xB72 | all | BSSS | 12043 | 47 | 0.004 |
| B78xB94 | all | BSSS | 11658 | 51 | 0.004 |
| B89xB84 | all | BSSS | 11517 | 53 | 0.005 |
| B11xB67 | cycle 0 | BSSS | 12195 | 45 | 0.004 |
| B17xB44 | cycle 0 | BSSS | 12230 | 54 | 0.004 |
| B39xB37 | cycle 0 | BSSS | 11485 | 126 | 0.011 |
| BB43xB69 | cycle 0 | BSSS | 12266 | 32 | 0.003 |
| B42xB54 | all | BSCB1 | 11260 | 38 | 0.003 |
| B90xB97 | all | BSCB1 | 6499 | 52 | 0.008 |
| B91xB95 | all | BSCB1 | 7830 | 56 | 0.007 |
| B99xB97 | all | BSCB1 | 6779 | 52 | 0.008 |
| BB42xB54 | cycle 0 | BSCB1 | 11260 | 57 | 0.005 |